\documentclass[aps,fleqn,superscriptaddress]{revtex4}
\usepackage{graphicx,color,natbib}
\usepackage{amsmath,amssymb,amsfonts}

\newcommand{\bse}{\begin{subequations}}
\newcommand{\ese}{\end{subequations}}
\newcommand{\be}{\begin{equation}}
\newcommand{\ee}{\end{equation}}
\newcommand{\bea}{\begin{eqnarray}}
\newcommand{\eea}{\end{eqnarray}}
\newcommand{\ba}{\begin{array}}
\newcommand{\ea}{\end{array}}

\input amssym.def
\input amssym.tex

\usepackage[colorlinks=true, linkcolor=blue, bookmarks=true]{hyperref}

\begin{document}
IPM/P-2017/086

\title{Dynamically probing strongly-coupled field theories with critical point}

\author{Hajar Ebrahim\footnote{hebrahim@ut.ac.ir}}
\affiliation{Department of Physics, University of Tehran, North Karegar Ave., Tehran 14395-547, Iran}
\affiliation{School of Physics, Institute for Research in Fundamental Sciences (IPM),
P.O.Box 19395-5531, Tehran, Iran}

\author{Mohammad Ali-Akbari\footnote{$\rm{m}_{-}$aliakbari@sbu.ac.ir}}
\affiliation{Department of Physics, Shahid Beheshti University G.C., Evin, Tehran 19839, Iran}

\begin{abstract}
The dependence of the rescaled equilibration time on different parameters of the field theories with a holographic dual has been investigated in this paper. We consider field theories with nonzero chemical potential at finite temperature which are dual to asymptotically AdS charged black holes. We examine a dynamical probe scalar operator where its dynamics is due to a time-dependent source, quantum quench, or out-of-equilibrium initial condition in field theory with fixed or varying temperature and chemical potential. We observe that the behavior of the scalar operator equilibration time with respect to temperature or chemical potential can not be predicted merely by field theory parameters and depends on how fast the energy is injected into the system. It is shown that in field theories with critical point the rescaled equilibration time is shorter for thermodynamically stable systems. We also observe that the rescaled equilibration time as one approaches the critical point enhances and acquires an infinite slope though its value remains finite. We show that for fast quenches, even though the system is far from equilibrium, the dynamical critical exponent is the same as the one reported for quasi-normal modes in the same background. However for slow quenches the dynamical critical exponent picks up a different value.

\end{abstract}

\maketitle

\tableofcontents

\section{Introduction and Results}
The physical phenomena around us in which we are mostly interested, involve systems out of equilibrium. Therefore, understanding the evolution of such systems towards  equilibrium is one of the important physics questions to be addressed. This kind of question can be approached from various directions, perturbative or non-perturbative. If the system is slightly out of equilibrium or near equilibrium methods like linear response theory or coarse-grained view of the system such as hydrodynamics can be applied \cite{Hubeny:2010ry, CasalderreySolana:2011us}. These methods deal with this problem in a perturbative way where the amplitude of the force which pushes the system away from equilibrium is small or equivalently, the gradient of the change in physical observables is not large. This question becomes harder to address if the system under study is strongly coupled and far from equilibrium. In such cases, the usual perturbative field theory methods are not applicable and one needs to deal with heavy numerical calculations. 

Lack of our knowledge to deal with strongly coupled systems has caused a lot of attention to be paid to gauge/gravity duality \cite{CasalderreySolana:2011us, Maldacena:1997re}. This conjecture duals a theory of classical gravity to a strongly coupled field theory in one less dimension. As a result different problems in strongly coupled gauge theories can be translated to dual ones in gravity theories, including far-from-equilibrium phenomena. For instance thermalization in field theory which generally means evolution of the state from zero temperature to a thermal state corresponds to black hole formation in the bulk gravity dual \cite{CasalderreySolana:2011us}. A lot of studies have been done in the literature to address different aspects of thermalization, for instance \cite{Chesler:2008hg}. 

An interesting question that needs to be addressed in far from equilibrium systems is how the system reaches its equilibrium state and how different parameters, for instance chemical potential and temperature, can affect this process. A far-from-equilibrium state in field theory can be produced by injecting energy into the system using an external source \cite{Buchel:2014gta}. One can also study the equilibration process, regardless of the presence of a source, by starting from an initial state which is far from equilibrium \cite{Heller:2013oxa}. Time-evolution of a far-from-equilibrium system can be probed using different observables such as local or non-local probes.    

In order to discuss what we are interested to study in this paper a couple of points should be mentioned. Let's consider the field theory temperature increases from zero to a final value. The time evolution can be probed by local and non-local operators in field theory. On the other hand these operators can have their own dynamics in the probe limit which does not affect the dynamics of the field theory. In the gauge/gravity framework these processes translate to black hole formation and to the dynamics of an external field in the probe limit in the bulk dual. The external field dynamics can involve adding a source or only considering out-of-equilibrium initial states. In other words, in addition to its own dynamics, the external field responds to the change in the temperature and other parameters in the field theory. 

In this paper we start with investigating the evolution of a scalar operator expectation value on a far-from-equilibrium or equilibrium state in the presence of nonzero temperature and chemical potential. Their gravity duals correspond to a Vaidya or a static charged black hole, respectively. In the Vaidya background the temperature and charge vary to reach their final values. The far-from-equilibrium state is assumed to be produced by an external probe scalar source or out-of-equilibrium initial condition. If the probe scalar source is nonzero it corresponds to quantum quench in field theory where a parameter such as mass or coupling varies time-dependently from zero to a final value. The probe scalar field in the bulk gravity with $m^2=-3$ that we consider in this paper is dual to a scalar operator with conformal dimension $\Delta=3$. The interested reader can refer to \cite{Buchel:2012gw} for more details. We would like to study the dependence of equilibration time on the parameters of the theory, chemical potential and temperature in field theory corresponding to the charge and mass in gravity dual. 

 Another interesting question to be addressed in this paper is to see whether the equilibration time knows about the phase structure of the field theory under study. In order to investigate this we consider a gravity background which its holographic dual is a field theory with a critical point. We will study the behavior of the equilibration time as the system moves towards the critical point in its phase diagram. 
 
 We briefly mention the main results of the paper as follows: 
 \begin{itemize}
 \item Studying the dependence of the rescaled equilibration time on the parameters of the theory shows that one can not report a general behavior for it with respect to temperature or the chemical potential. In cases where the probe has its own dynamics the raise or fall in the rescaled equilibration time with respect to field theory parameters depends on the speed of energy injection.
 \item   Interestingly we observe that the rescaled equilibration time, in the field theories we consider here, can be fitted with second order polynomial in dimensionless parameter $\frac{\mu}{T}$ for various choices of field theory parameters. 
 \item We observe that in field theories with the critical point the rescaled equilibration time of the scalar operator knows whether the field theory parameter $\frac{\mu}{T}$ belongs to thermodynamically stable configuration. We show that shorter rescaled equilibration time belongs to stable branch of the parameter $\frac{\mu}{T}$. 
 \item The rescaled equilibration time remains finite as we move towards the critical point but its slope diverges at the critical point and behaves like $((\frac{\mu}{T})_*-\frac{\mu}{T})^{-\theta}$ where $(\frac{\mu}{T})_*$ denotes the value of $\frac{\mu}{T}$ at the critical point. We observe that $\theta$ depends on whether the quench is fast or slow. We define $\theta$ as the dynamical critical exponent following \cite{Finazzo:2016psx}.
\item A very intriguing observation is that for fast quenches, even though the system is far from equilibrium, the dynamical critical exponent is very close to $0.5$ which is the same as the dynamical critical exponent obtained from the behavior of quasi-normal modes near the critical point in such backgrounds \cite{Finazzo:2016psx}.  

 \end{itemize}

\section{The Charged Black Hole Background}
Heavy ion experiments done at LHC, RHIC and future colliders are good examples of producing a medium, called quark-gluon plasma, that is strongly coupled and out of equilibrium \cite{CasalderreySolana:2011us}. Comparing experimental observations and hydrodynamic simulations suggests that viscosity normalized by entropy density is very small for such plasma and therefore indicates that the plasma is strongly coupled. Another outcome of this comparison is the fact that the plasma thermalizes very fast, by which we mean hydrodynamic equations can describe the dynamic of the system very soon after the collision. Before hydrodynamic equations are applied, the plasma is out of equilibrium and this is the stage we are trying to model and study in this paper. Due to the collision of heavy nuclei in theses experiments the net baryon number is nonzero which corresponds to a nonzero chemical potential although it's very small. Thus in the QCD phase space the plasma is situated in the crossover region, near the temperature axis. 

As explained before in order to study such system we use gauge/gravity duality. Since we are interested in studying the effect of field theory parameters such as chemical potential on equilibration in out-of-equilibrium systems we will consider the background with a $U(1)$ gauge field.  Due to the presence of a gauge field in the bulk the chemical potential is nonzero. 

The background metric that we study is a solution to the following action
\be
S=\frac{1}{16 \pi} \int d^5x \sqrt{-g}\left( {\cal{R}}-\frac{4}{3} (\nabla \Phi)^2 - V(\Phi) - e^{- \frac{4 \alpha}{3} \Phi} F_{\mu\nu}F^{\mu\nu}\right),
\ee
where ${\cal{R}}$, $F_{\mu\nu}$ and $\Phi$ are the Ricci scalar, gauge field strength and the scalar field, respectively. $V(\Phi)$ is the scalar potential which one can find its details in \cite{Zhang:2015dia}. $\alpha$ determines the coupling constant between the gauge and scalar field on the gravity side. If $\alpha=0$ we recover the known Einstein-Maxwell-scalar theory. If $\alpha=1$ the last term in the action gives the dilaton-Maxwell coupling that appears in the low energy string action in Einstein's frame. The metric in the solution to this action is
\be
ds^2 = -N(z) f(z) dv^2 - \frac{2}{z^2} \sqrt{\frac{N(z)}{1+b^2 z^2}} dv dz+\frac{1+b^2 z^2}{z^2} g(z) d{\vec {x}}^2,
\label{back}
\ee
where $z$ is the radial coordinate and 
\bea
f(z) &=& \frac{1+b^2 z^2}{z^2} \Gamma^{2 \gamma} -m \frac{z^2}{1+b^2 z^2} \Gamma^{1-\gamma},\cr
N(z) &=& \Gamma^{-\gamma},~g(z)=\Gamma^{\gamma},~\Gamma(z)=1-\frac{b^2 z^2}{1+b^2 z^2},~\gamma=\frac{\alpha^2}{2+\alpha^2}.
\eea 
Note that $m$ is the mass of the black hole.  The relation between the charge of the black hole, $q$, and its mass is
\be
q=\sqrt{\frac{6 m}{2+\alpha^2}} b.
\ee
This solution is asymptotically AdS$_5$ and its boundary is located at $z=0$. The field theory lives on $(t,\vec {x})$ where $t$ is $v$ at the boundary. We have set the AdS radius to one. The full solution to the above action involves nontrivial scalar and gauge field which, for more information, the reader can consult \cite{Zhang:2015dia}. The chemical potential in the field theory, due to the gauge field in the bulk, becomes \cite{Zhang:2015dia}
\be
\label{mu}
\mu =\frac{b \sqrt{3 m}}{\sqrt{2 \left(\alpha ^2+2\right)} \left(b^2+\frac{1}{z_h^2}\right)},
\ee 
in AdS radius unit. $z_h$ is the horizon radius, the largest root of the equation $f(z_h)=0$. The Hawking temperature of the black hole is
\be
\label{t}
T=\frac{b \Gamma (z_h)^{\frac{3 \gamma }{2}-1}}{4 \pi \sqrt{1-\Gamma (z_h)}} \left(2 (3 \gamma -1)-3 (2 \gamma -2) \Gamma (z_h)\right).  
\ee  
This background is parametrized by three parameters $m,q,b$, where the two independent ones in field theory dual are temperature and chemical potential. Since we are interested in studying the thermalization process in field theory, we need to consider the background to be of Vaidya type. This type of the background can be built by replacing $f(z)$ with $f(v,z)$ which translates into exchanging $m$ with $m P(v)$ and $q$ with $q \sqrt{P(v)}$. $P(v)$ is the time-dependent function which we choose to be
\be
P(v) = \frac{1}{2} \left( 1+\tanh \frac{v}{\beta_{BH}} \right),
\ee
where $\beta_{BH}$ specifies at what speed the background changes.

In order to explain what the parameter $\alpha$ represents in the field theory we look at the solution to the scalar field $\Phi$
\be
\Phi(z) =\frac{3}{4} \sqrt{2 \gamma (1+ \gamma-2 \gamma)}~\ln \Gamma(z).
\ee
It is known that, according to gauge/gravity duality, the equation of motion for the scalar field in the bulk with appropriate boundary conditions corresponds to $\beta$-function of the coupling in field theory \cite{link}. Therefore, regarding the above solution for $\Phi$, $\alpha$ represents different coupling flow directions to the UV fixed point in field theory.

\section{Dynamical probe and equilibration time}
As was mentioned before we would like to study the process of equilibration and its response to the parameters introduced in the physical system. The origin of this desire comes from the fact that the quark-gluon plasma produced in heavy ion collisions is out of equilibrium at the very early stages of its evolution. We try to model a system which in some sense can give us some information about the equilibration and how it is affected by parameters relevant to quark-gluon plasma. 

\subsection{Set-Up}
We consider a probe scalar field in the background, introduced in the previous section, which has its own dynamics and evolves in time. According to the gauge/gravity duality, the near boundary expansion of the scalar field in the bulk gives the source and expectation value of the corresponding scalar operator in the field theory where the expectation value is calculated on an arbitrary state. Since we are interested in the equilibration process, we study the response of a dynamical scalar operator on an equilibrium or out-of-equilibrium state corresponding to charged static or Vaidya black hole background in the gravity dual. The non-trivial response of the scalar operator can be resulted from various time-dependent initial conditions or a time-dependent source. 

A scalar field is added to the background in the probe limit and to study its dynamic we solve the corresponding Klein-Gordon equation. We have assumed the mass of the scalar field is $m^2=-3$, corresponding to an operator with mass dimension $\Delta=3$ in field theory. It needs to be emphasized that this scalar field $\phi(v,z)$ should not be confused with the scalar field $\Phi(z)$ in the background. The near boundary expansion of the scalar field can be derived in the form
\bea
\phi(v,z)= z\phi_s(v)+z^2 \phi_s'(v)&+&z^3 \left(\frac{1}{2} \log(z) \phi_s''(v)+\phi_r(v)\right)\cr
&+&z^4 \left(\frac{6 m \phi_r'(v)-2 m \phi_s'''(v)-q^2 \phi_s'(v)}{6 m}+\frac{1}{2} \log(z) \phi_s'''(v )\right)+ \mathcal{O}(z^5),
\eea
where $\phi_s(v)$ and $\phi_r(v)$ are the source and response in field theory, respectively, obtained from the asymptotic expansion of the scalar field in the bulk. The derivatives of these functions are taken with respect to $v$. We have chosen the time-dependent source in field theory to be
\be
\label{12}
\phi_s(v)=\frac{1}{2} \left( 1+\tanh \frac{v}{\beta_{\phi}} \right),
\ee
where $\beta_{\phi}$ identifies the speed at which the energy is injected into the field theory. In order to solve the Klein-Gordon equation for the scalar field in the bulk we have to impose appropriate boundary and initial conditions\footnote{If there is no source or the initial condition is zero for the scalar field, $\phi(v,z)=0$ is the solution to the Klein-Gordon equation at all times.}. It is more practical to use the following notation for the scalar field
\be
\phi(v,z)= z\phi_s(v)+z^2 \phi_s'(v)+z^3 \frac{1}{2} \log(z) \phi_s''(v)+ z {\tilde{\phi}}_r(v,z).
\ee
Therefore the appropriate boundary conditions imposed on the scalar field are
\be
 {\tilde{\phi}}_r(v,0)=0,~\partial_z{\tilde{\phi}}_r(v,0)=0.
 \ee
 To set the proper initial conditions we have to distinguish between zero and nonzero sources. For the case where the source is not zero, the initial condition is
 \be 
 \label{inisource}
 {\tilde{\phi}}_r(v_0,z)=0,
 \ee
 where $v_0$ is the initial time. We choose $v_0=-20$.  When the source is zero we choose the initial condition to be
 \be
 \label{initial}
{\tilde{\phi}}_r(v_0,z)=z^4.
\ee
We can generally have different initial conditions as long as they satisfy the near boundary expansion of the scalar field. Solving the equation of motion of the scalar field we will be able to obtain ${\tilde{\phi}}_r(t=v|_{z=0})$. This in fact is the response function on the boundary. In the field theory the expectation value of the scalar operator dual to the bulk scalar field is proportional to $\langle \mathcal{O}(t) \rangle \approx \partial_z^2 {\tilde{\phi}}_r(t,z)_{z=0}$. Since ${\tilde{\phi}}_r(t)$ evolves with time in the non-equilibrium situation we define a function as 
\be
\label{epsilon}
\epsilon(t)=\bigg{|}\frac{\langle \mathcal{O}(t) \rangle-\langle \mathcal{O}(t=\infty) \rangle}{\langle \mathcal{O}(t=\infty) \rangle}\bigg{|},
\ee
and $t_{eq}$ is defined as the the time $\epsilon(t_{eq})<5\times 10^{-3}$ and $\epsilon(t)$ stays below this value afterwards. 

\begin{figure}
\centering
\includegraphics[width=70mm]{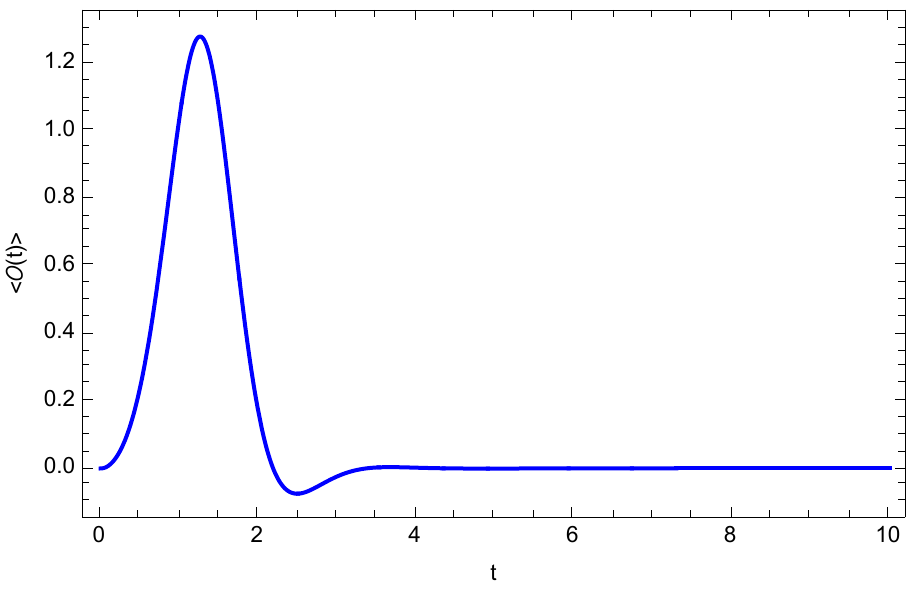}
\includegraphics[width=70mm]{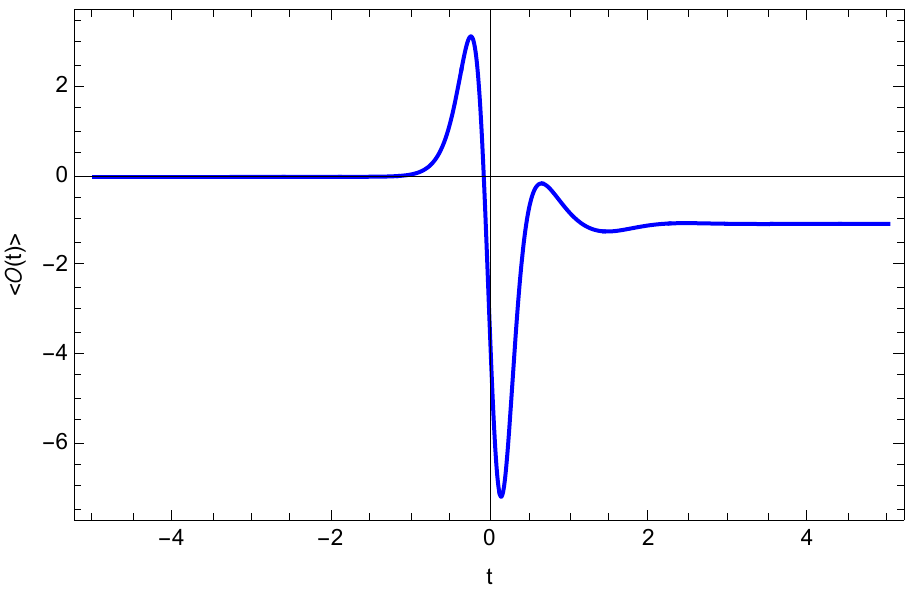}
\caption{time evolution of the expectation value of the scalar operator in field theory with initial condition (left) set as $z^4$ and in the presence of a source (right) for  $\frac{\mu }{T} = 1.26908$ and $\alpha=1.5$.} \label{scalar12}
\end{figure}

We conclude this section with two important comments. First, we expect various initial conditions do not alter thermalization time substantially and keep the general behavior unchanged. It can be checked by numerical results. Also the results of the papers \cite{Shahkarami:2017fxc, Ali-Akbari:2016sms} approve our expectation. Second, when the scalar source is zero and we are dealing with nonzero initial conditions the final equilibrium scalar response is zero and hence the definition of $\epsilon(t)$ in equation \eqref{epsilon} reduces to  $\epsilon(t)=|\langle \mathcal{O}(t) \rangle|$. In order to see this more clearly we have plotted the time evolution of the scalar operator in field theory in figure \ref{scalar12}. As one can see in the right plot the expectation value of the scalar operator reaches a nonzero constant value after a while. 

\begin{figure}
\centering
\includegraphics[width=100mm]{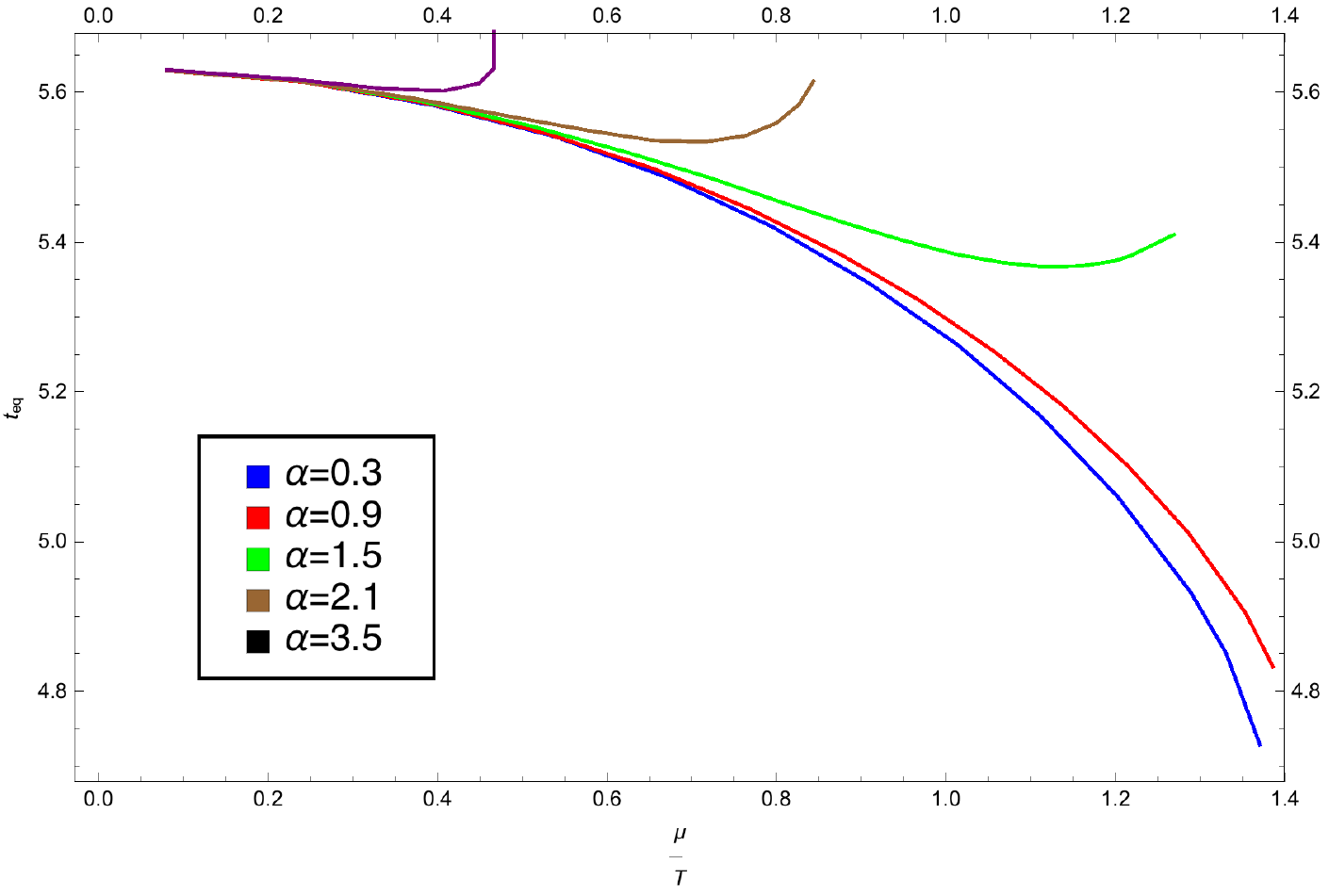}
\caption{equilibration of scalar field with initial condition set as $z^4$ for different values of $\alpha$, for the static charged background metric \eqref{back}} \label{initalpha}
\end{figure} 

\subsection{Numerical Results}
In order to study the effect of different physical parameters on the equilibration time, we start with the assumption that the temperature is fixed and the chemical potential varies. Also the coupling flow directions to the UV fixed point changes with the value of $\alpha$. We choose the temperature to be $T=0.37$ and plot the equilibration time with respect to $\frac{\mu}{T}$ for different values of $\alpha$, figure \ref{initalpha}. One should note that all relevant parameters are measured in units of AdS radius equal to one. It should be emphasized that in this figure we start from the initial out-of-equilibrium states with the initial condition set as in equation \eqref{initial}.  It can be realized from the figure for small values of $\mu$, the equilibration time is independent of $\alpha$ as all different curves coincide. For the range of $\mu$ we considered in this paper it seems that the equilibration time is a decreasing function in $\mu$ for small values of $\alpha$. As $\alpha$ reaches the values larger than one, a minimum occurs in the curves and the behavior of the equilibration time reverses. The equilibration time increases afterwards quite rapidly as $\mu$ raises. These results confirms the results obtained in \cite{Zhang:2015dia} for different $\alpha$s. It should be mentioned that for $0 < \alpha < 1$, $\frac{\mu}{T}$ is a monotonically increasing function of $q$. As the values of $b$ and $q$ varies from zero to a maximum value, $\frac{\mu}{T}$ spans the whole range of values from zero to infinity. But for $\alpha \geq 1$, $\frac{\mu}{T}$ is bounded and is no longer a monotonically increasing function of $q$. It can be seen from figure \ref{initalpha} that for $\alpha \geq 1$ the equilibration time reaches a minimal value and the behavior reverses and starts increasing afterwards in a multivalued manner which we have not shown here. 

We have also checked the conclusion discussed in previous paragraph for the case where there is a nonzero source for the scalar operator in the boundary. As $\alpha$ increases there will be a minimum value in the equilibration time and it increases afterwards and the behavior observed in figure \ref{initalpha} persists here, too. 

\begin{figure}
\centering
\includegraphics[width=85mm]{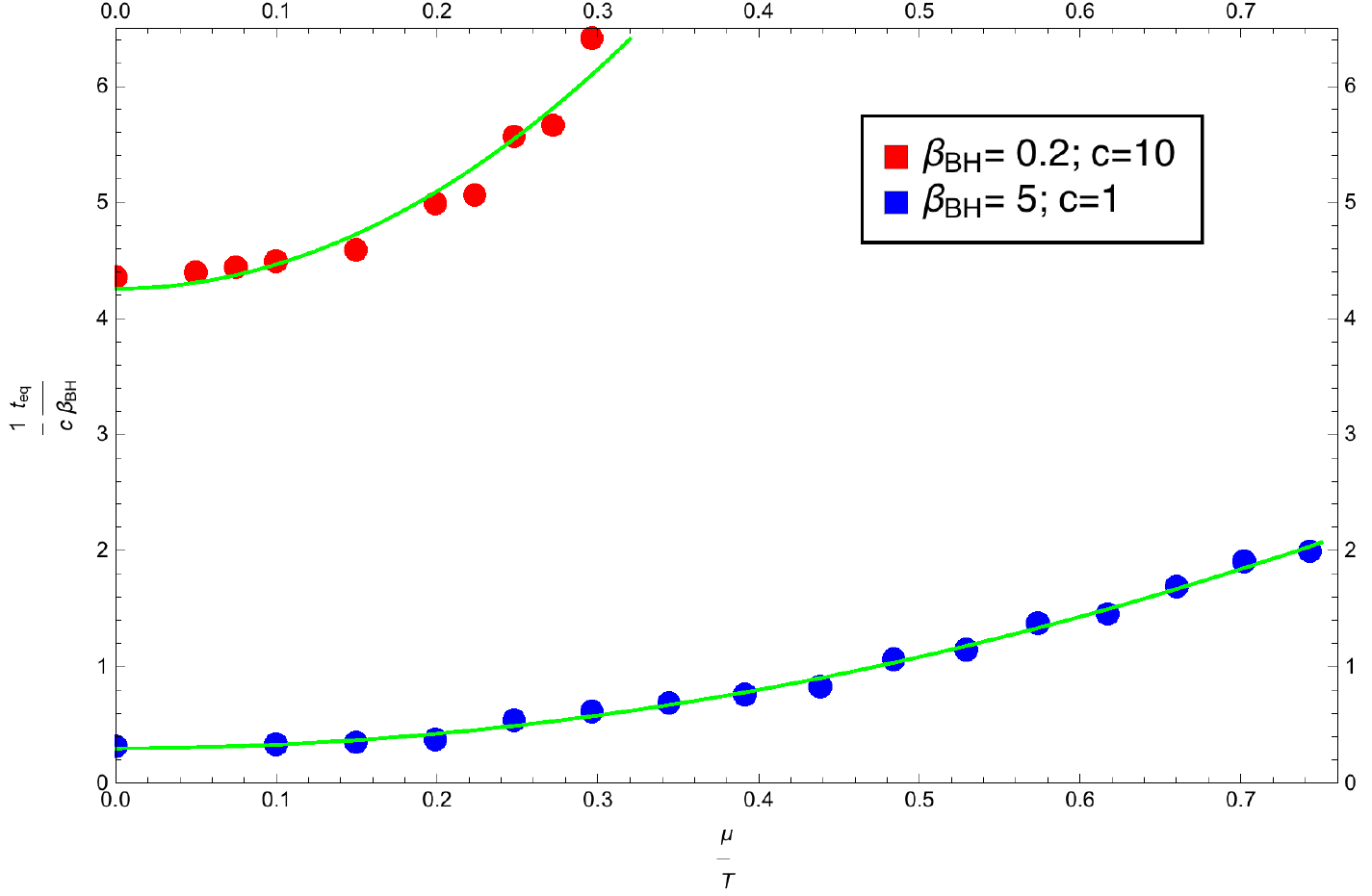}
\includegraphics[width=85mm]{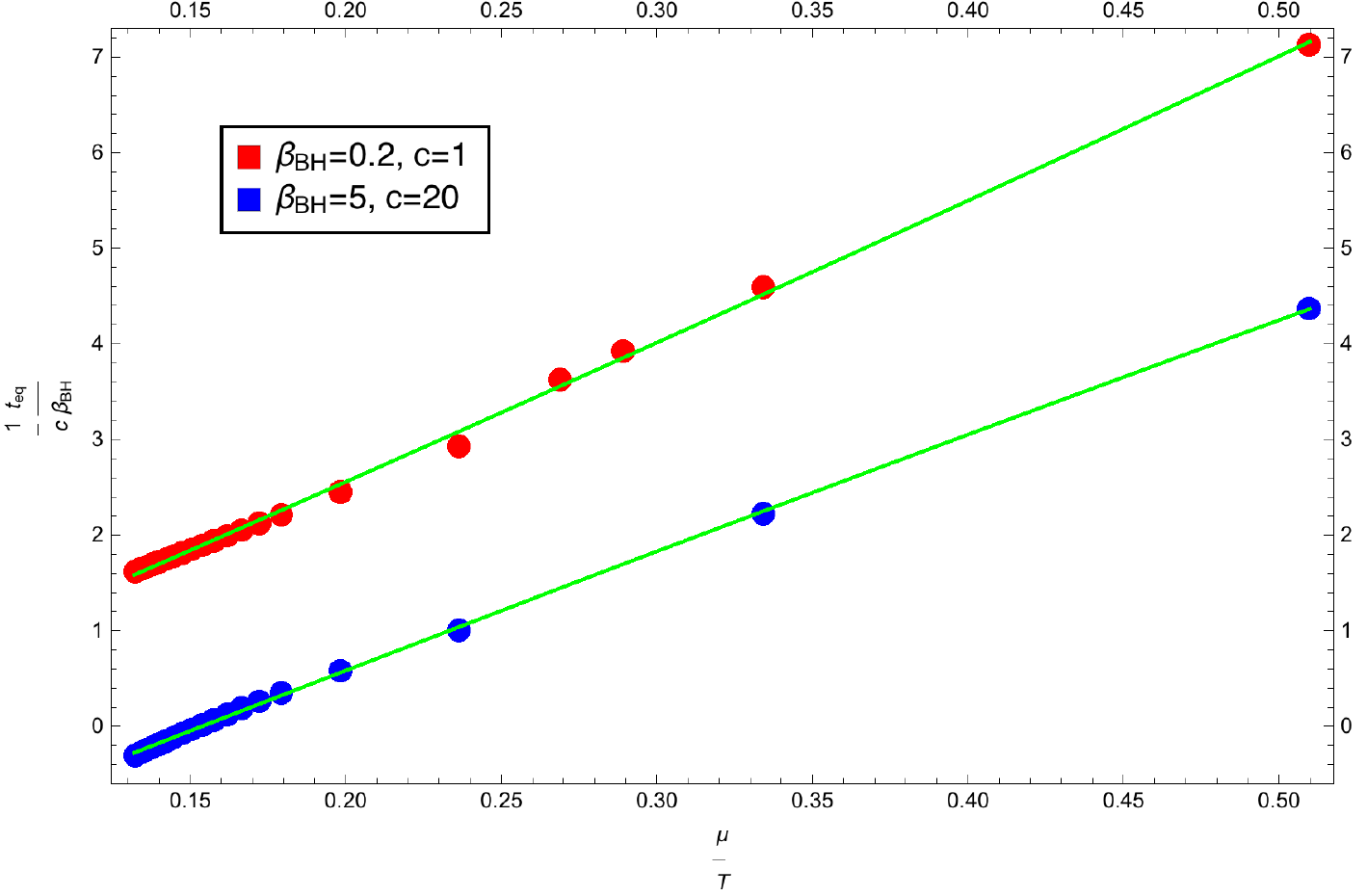}
\caption{The rescaled equilibration time with respect to $\frac{\mu}{T}$ in the field theory dual to the Vaidya charged black hole has been plotted for various final values of $\mu$ at fixed final temperature, $T=0.37$, (left) and various final values of $T$ at fixed final chemical potential, $\mu=0.05$, (right). The scalar source is zero and the initial condition is set as $z^4$. Please note that the oscillation of the points in the left plot is due to the choice in $\epsilon<5\times 10^{-3}$ and the oscillations disappear if we choose other values.} \label{initialvaidya}
\end{figure}

As was mentioned before we are interested in studying the equilibration process in field theory. We will consider an out-of-equilibrium state which is produced by time-dependent temperature in field theory and is dual to Vaidya background in the bulk. The scalar probe has dynamics due to its out-of-equilibrium initial condition and the Vaidya background. We have studied the equilibration time in two cases where the final temperature is kept fixed, left plot of figure \ref{initialvaidya}, or the final chemical potential is fixed, right plot of figure \ref{initialvaidya}. Note that in these plots the equilibration time is scaled with the parameter $\beta_{BH}$ which makes the rescaled equilibration time dimensionless and presents how fast or slow the temperature and chemical potential varies in field theory. In general $\beta\gg 1$ ($\beta\ll 1$) represents slow (fast) quench. Also note that the numerical coefficient $c$ has been given the numbers specified in plots in order to be able to show all the data points in one figure. All the points in this figure can be fitted with {\it{second order polynomials}} of $\frac{\mu}{T}$ and in the range of $\frac{\mu}{T}$ considered the rescaled equilibration time increases with $\frac{\mu}{T}$. A difference is where the temperature changes not substantially fast, blue dots in right plot of this figure. Although all the other plots are concave upward, the mentioned plot is concave downward. Also another difference that can be seen is that although the points are all fitted with quadratic equations in $\frac{\mu}{T}$ but when the chemical potential is kept fixed the change in the rescaled equilibration time is almost linear in contrast to when the temperature remains unchanged especially for very rapid change due to $\beta_{BH}=0.2$. An important point should be highlighted here that having a scalar field in the bulk with $m^2=-3$ corresponds to adding a fermionic mass operator in the boundary theory with $\Delta=3$. Therefore a new scale in field theory will be introduced. As a result the behavior of the equilibration time with respect to $\frac{\mu}{T}$ is not necessarily similar for fixed $\mu$ and fixed $T$ as we will see in this figure.

\begin{figure}
\includegraphics[width=85mm]{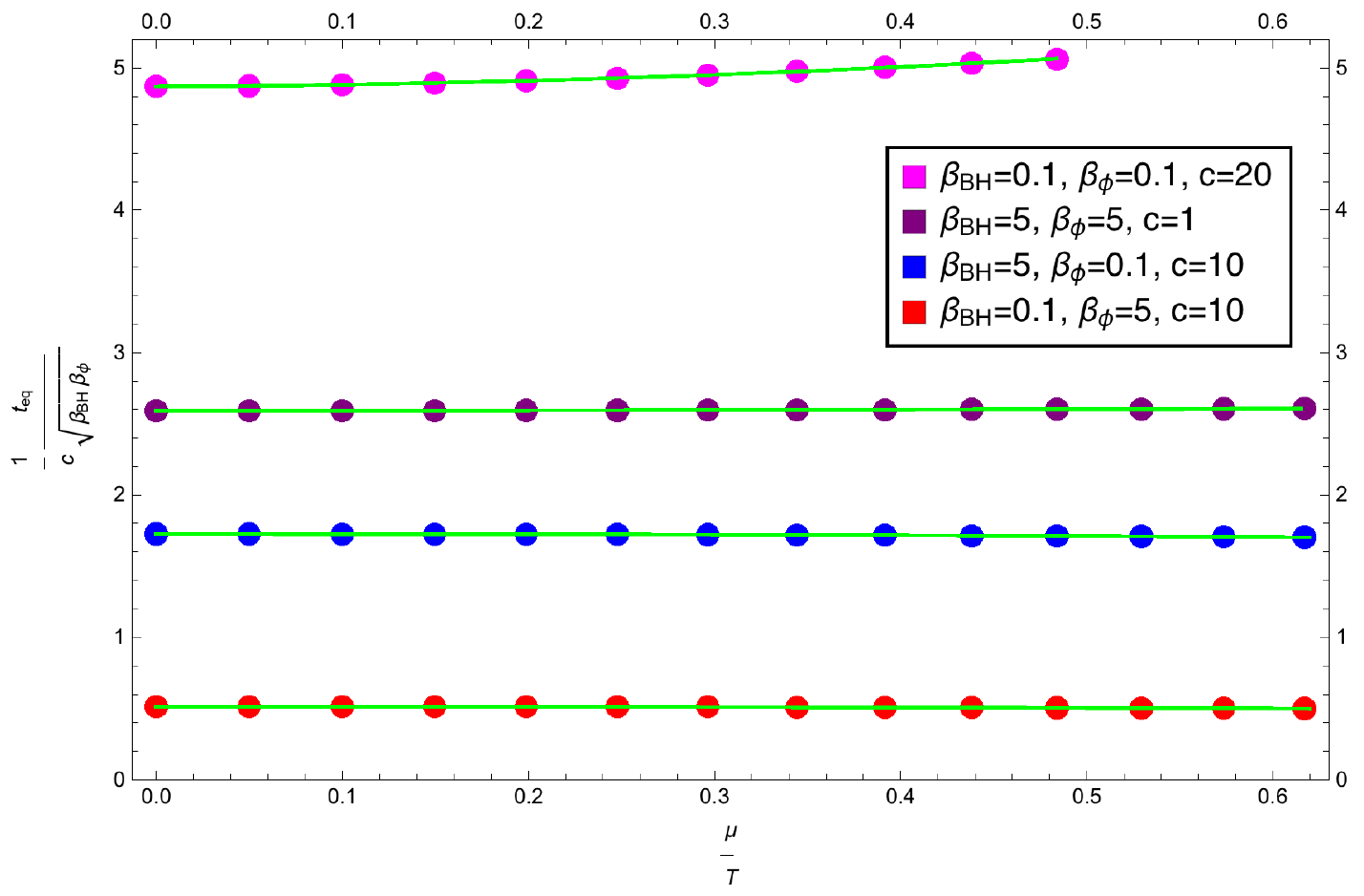}
\includegraphics[width=85mm]{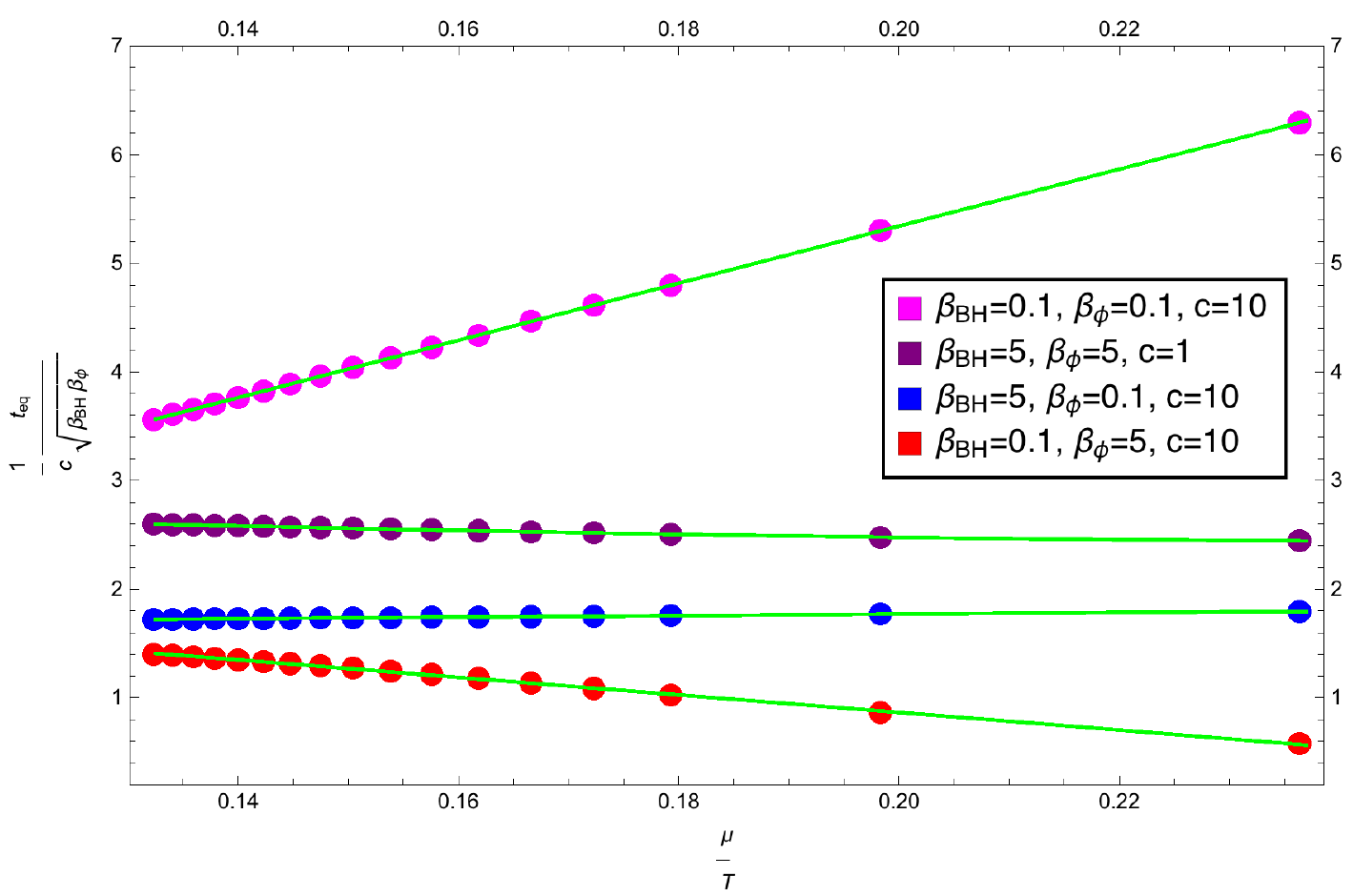}
\caption{The rescaled equilibration time with respect to $\frac{\mu}{T}$ in the field theory dual to the Vaidya charged black hole has been plotted for various final values of $\mu$ at fixed final temperature, $T=0.37$, (left) and various final values of $T$ at fixed final chemical potential, $\mu=0.05$, (right). The scalar source is set as \eqref{12}.} \label{slow3}
\end{figure}

\begin{figure}
\includegraphics[width=87mm]{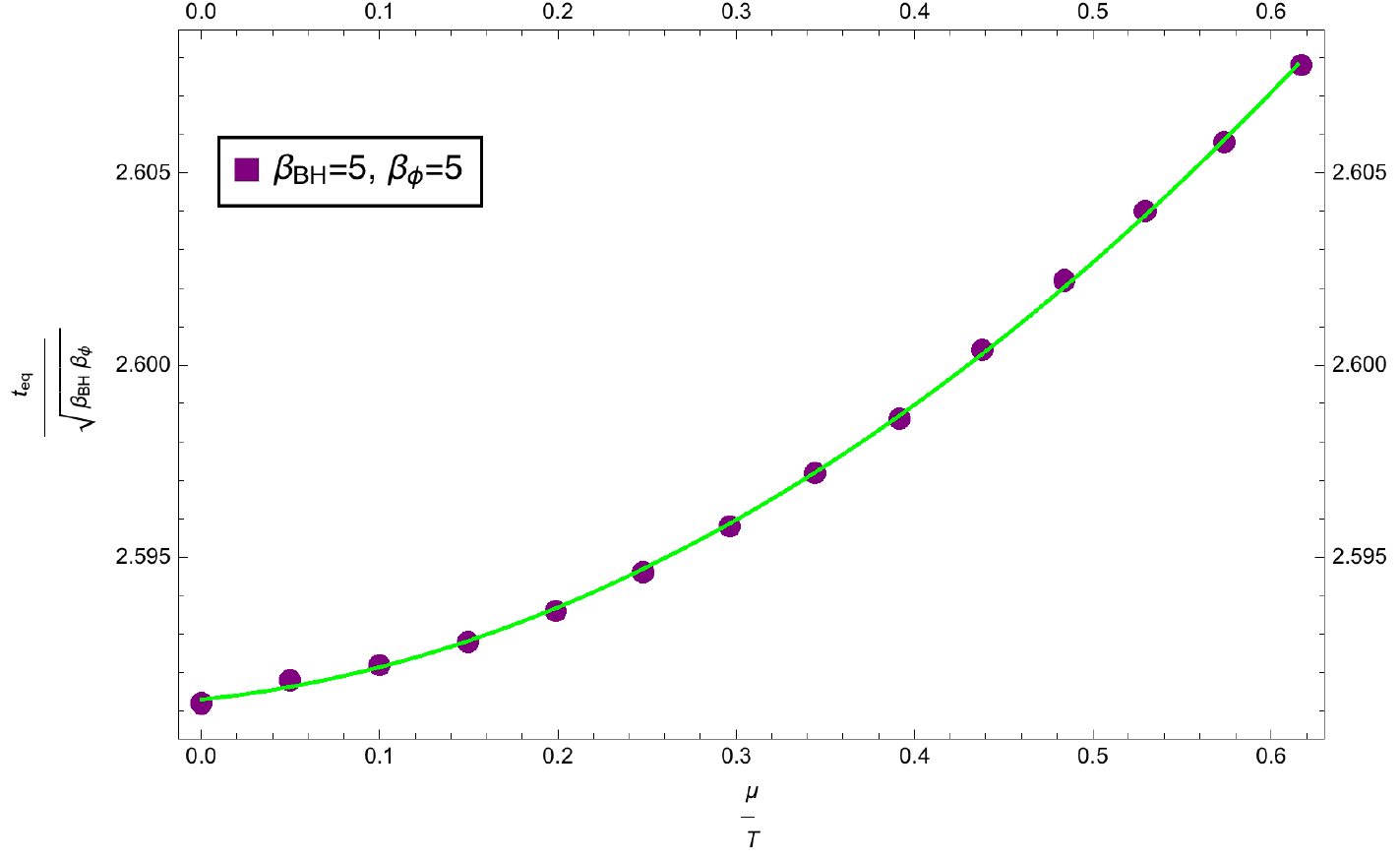}
\includegraphics[width=85mm]{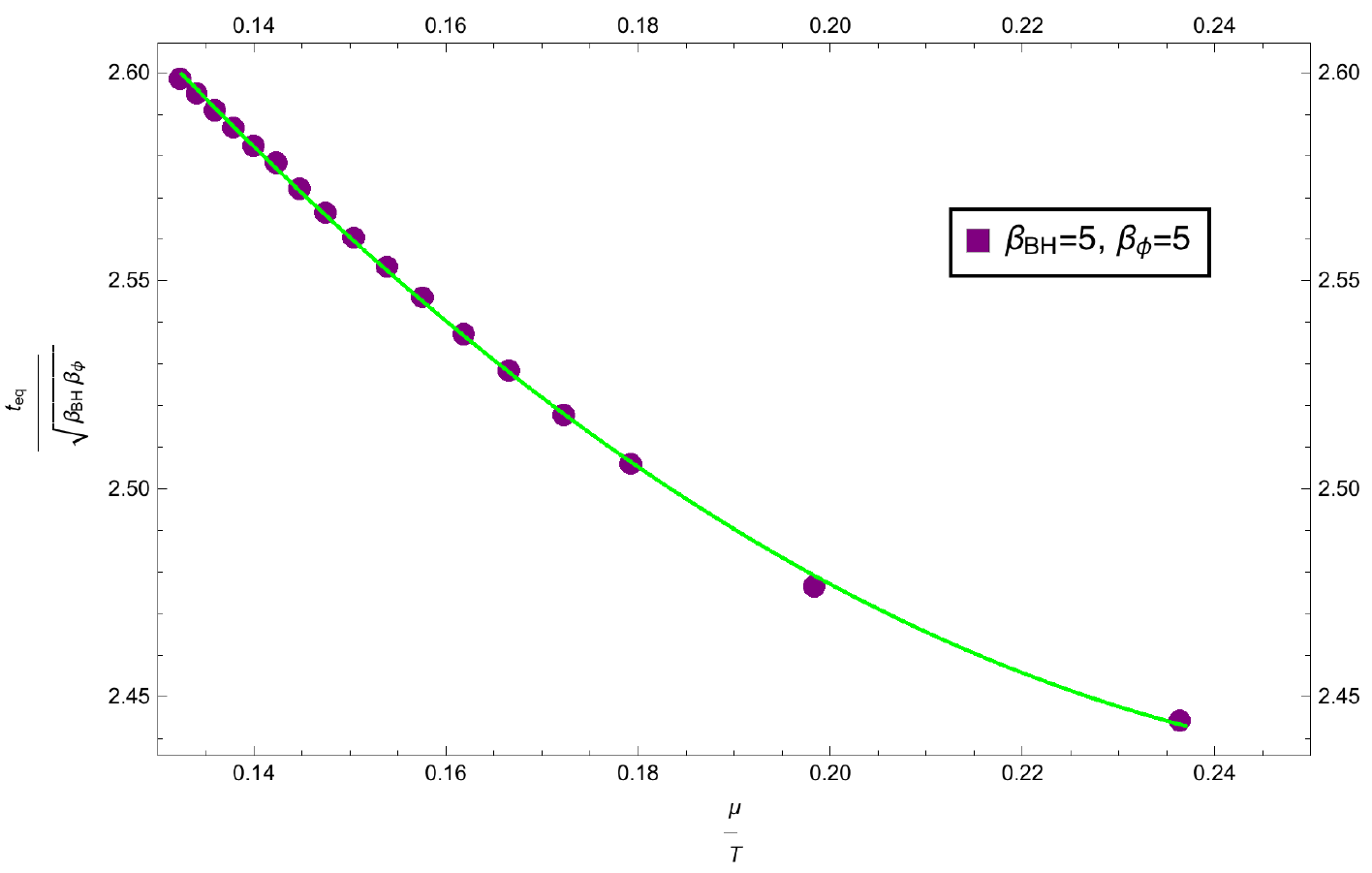}
\caption{two of the curves in figure \ref{slow3}} \label{compare}
\end{figure}  

Next we have studied the rescaled equilibration time for the case where there is a nonzero source for the scalar field and the background is Vaidya metric. Therefore we can investigate how system equilibrates in rapid and slow energy injections defined by $\beta_{\phi}$ and $\beta_{BH}$. We have plotted the results in figure \ref{slow3} for the cases where final $\mu$ or $T$ is kept fixed. The left figure shows the dependence of rescaled equilibration time on $\mu$ for different combinations of $\beta$s. It can be obviously concluded that the dependence of rescaled equilibration time on chemical potential is rather mild. But if we check it more closely we can extract that for both energy injections being rapid or both slow the rescaled equilibration time increases as the chemical potential is raised. But if the injections are one rapid and one slow the behavior reverses and the slope in rescaled equilibration time is negative. The points to a good approximation fit with the polynomial curves (green curves in the figure) of second order in ${\frac{\mu}{T}}$. In the right plot the chemical potential is kept fixed and the temperature varies. As it can be clearly seen varying the temperature can affect the rescaled equilibration time more substantially compared to varying the chemical potential in the left figure. Similar to the right figure we can see that there are both decreasing and increasing curves of $\frac{\mu}{T}$ but not with the same combinations of $\beta$s. In both plots, right and left, the maximum change in the rescaled equilibration time happens for very rapid injections of energies, both $\beta$s small.  

From figure \ref{slow3} one can conclude that the values of $\beta$s can not completely determine the increasing or decreasing behavior of rescaled equilibration time with respect to varying $\frac{\mu}{T}$. For example in figure \ref{compare} we have plotted two of the curves with the same combination of $\beta$s in the previous figure, separately and one can clearly see that based on which parameter varies, $\mu$ or $T$, we can have both behaviors.  

The results obtained so far can be compared with the previous works on this topic where it is claimed that basically with the raise in chemical potential the rescaled equilibration time increases \cite{Galante:2012pv}. We should emphasize that in those works the probe is a local or non-local probe which does not have dynamics on its own. From the discussions, up to here, we can conclude that when the probe dynamically evolves on its own, due to an external source or various initial out-of-equilibrium conditions, one can not generally predict the behavior of the rescaled equilibration time with respect to the temperature or chemical potential. In other words the raise or fall in the rescaled equilibration time depends on the details of the problem considered. Another observation general to all plots is that the rescaled equilibration time behaves like a second order polynomial in $\frac{\mu}{T}$. The green curves are the polynomials fitted with the data points. 

\section{Probing the critical point}
So far we have studied how the behavior of equilibration time depends on $\alpha$, chemical potential and temperature in field theory. An interesting question one might ask is to see whether the response in field theory to an external source understands about the phase structure of the theory. To investigate this problem we consider the background with $\alpha=2$. With some field redefinition this is in fact the holographic model of the QCD critical point constructed in \cite{DeWolfe:2010he}. Time-dependent perturbations to this black hole has been studied in \cite{Finazzo:2016psx, Rougemont:2015wca}. Some of other papers in this context are \cite{Critelli:2017euk}.

\begin{figure}
\includegraphics[width=81mm]{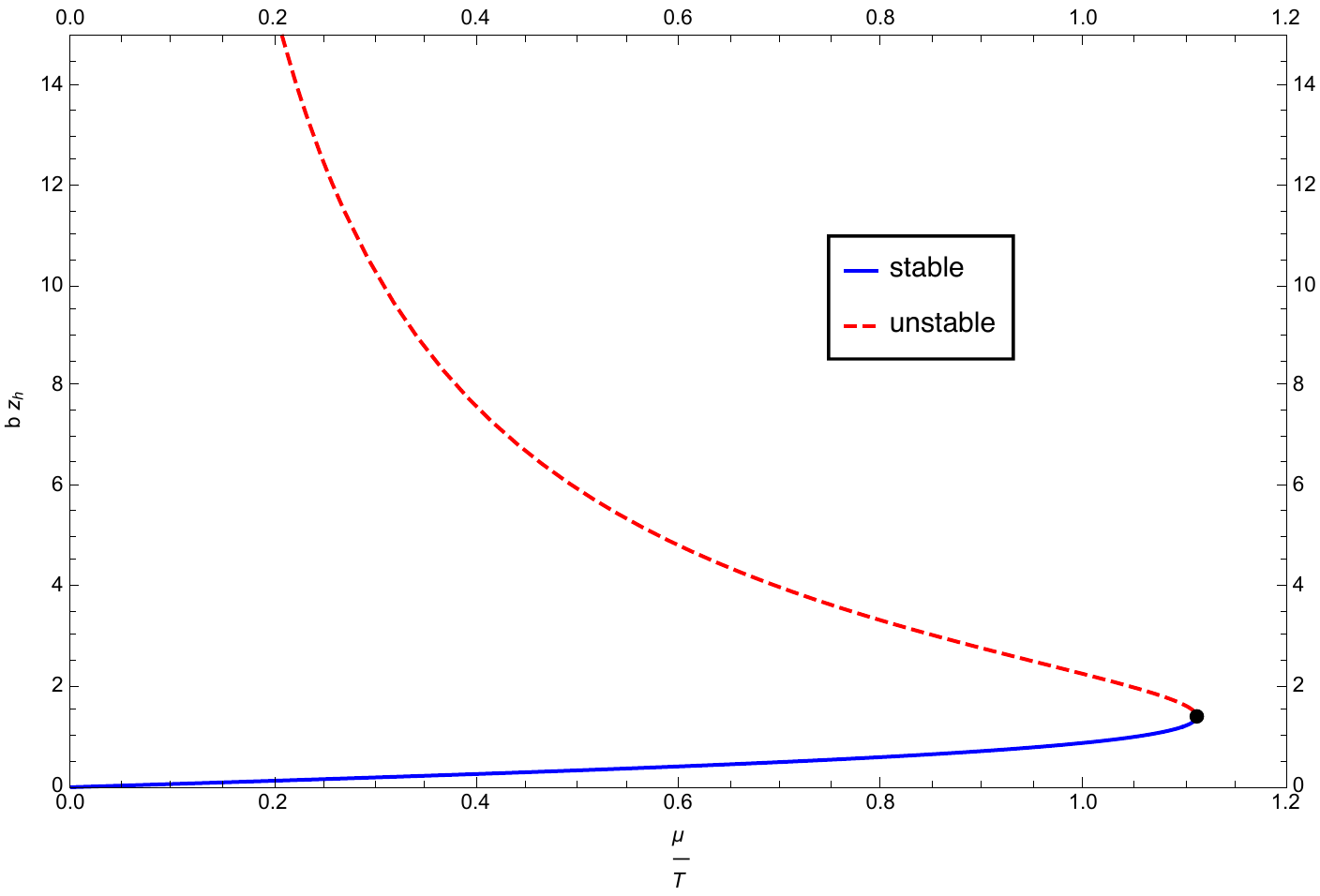}
\includegraphics[width=80mm]{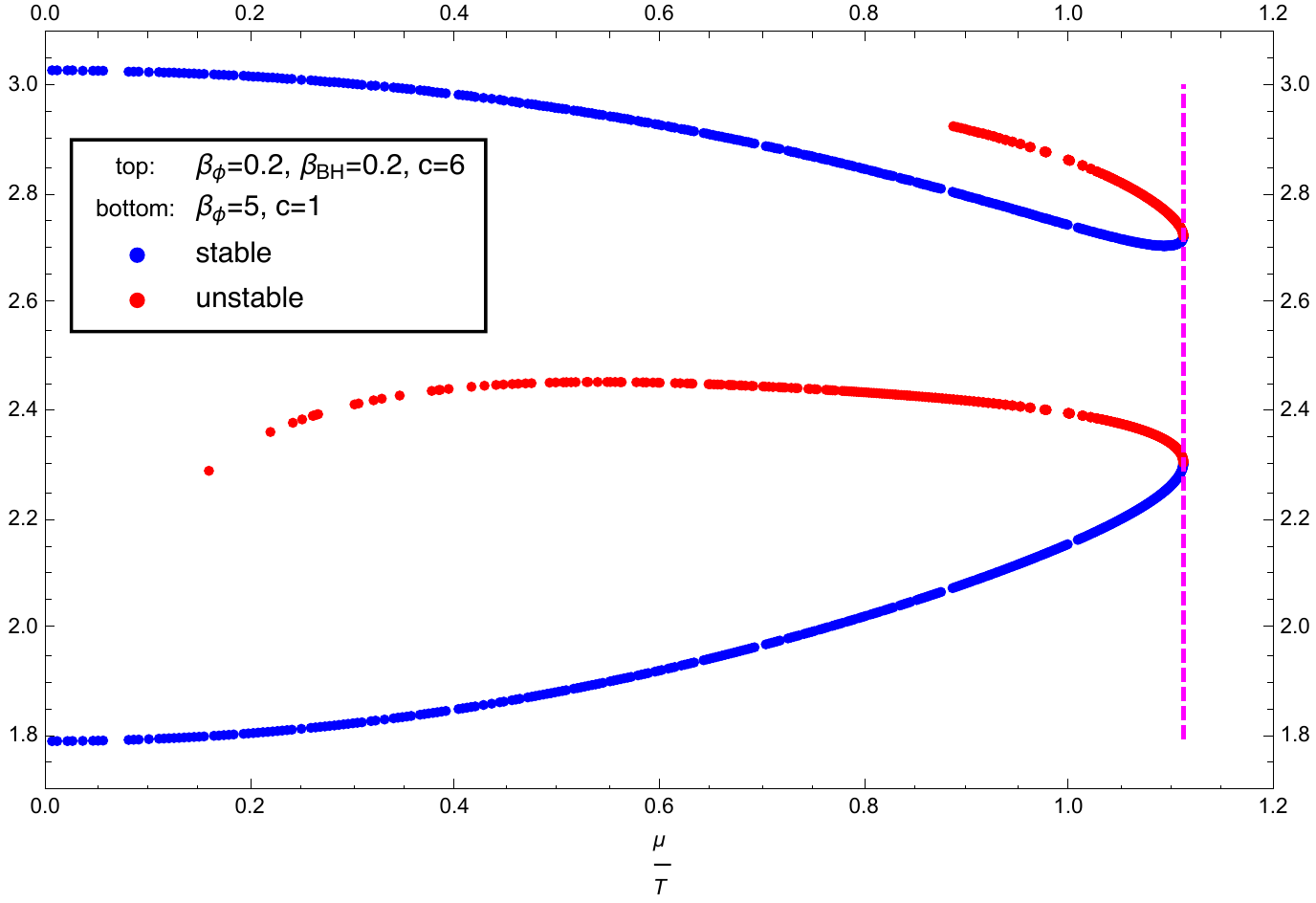}
\caption{Left: This figure shows the dependence of $b z_h$ on $\frac{\mu}{T}$. For each value of $\frac{\mu}{T}$ there exists two distinct values of $b z_h$. The black dot shows the critical point at $(\frac{\pi}{2 \sqrt{2}},\sqrt{2})$. Right: This figure shows the rescaled equilibration times, $t_{eqs}=\frac{t_{eq}}{c \sqrt{\beta_{\phi} \beta_{BH}}}$ and $\frac{t_{eq}}{c \beta_{\phi}}$ for top and bottom plot, respectively, with respect to $\frac{\mu}{T}$ for the fixed final temperature, $T=0.37$. The magenta dashed line corresponds to the critical point  which is at $(\frac{\mu}{T})_*=1.1107$.} \label{critical}
\end{figure}

The phase structure of a system can be studied using the behavior of heat capacity at fixed chemical potential and other appropriate thermodynamic quantities. The divergence in some of these quantities indicates the presence of a critical point. More details of the behavior of thermodynamic quantities near the critical point in the field theory dual to this background can be found in \cite{Finazzo:2016psx} and \cite{DeWolfe:2010he}. Since we use a different notation in our paper we review some facts about the phase structure of the solution. To do so we obtain the dependence of the black hole parameters on $\frac{\mu}{T}$, the parameter in field theory dual. From the relations \eqref{mu} and \eqref{t} one can see that  
\be
\label{bz}
b z_h=\frac{1\pm \sqrt{1-\frac{8 \mu^2}{\pi^2 T^2}}}{\frac{2 \mu}{\pi T}},
\ee
where
\be
z_h=\sqrt{\frac{b^2+\sqrt{b^4+4 m}}{2 m}}.
\ee
It's clear that each value of $\frac{\mu}{T}$ corresponds to two distinct values of $b z_h$ which parametrizes stable and unstable branches of solutions as shown in figure \ref{critical}, left. This conclusion indicates the existence of phase transition in field theory. One branch of the plot corresponds to thermodynamically stable configurations and the other branch, unstable. This can be checked using the Jacobian, $\mathcal{J}=\frac{\partial (s,\rho)}{\partial (T,\mu)}$ where $s$ and $\rho$ are entropy and charge density, respectively
\bea
s&\propto& \frac{T^3 (1+b^2 z_h^2)^2}{(2+b^2 z_h^2)^3},\\
\rho &\propto& \frac{\mu}{T} (2+b^2 z_h^2) \sqrt{1+b^2 z_h^2}.
\eea
If the Jacobbian is positive for a set of parameters then the systems corresponding to those are thermodynamically stable \cite{Finazzo:2016psx}. The upper(lower) sign in \eqref{bz} correspond to thermodynamically unstable(stable) configurations. The maximum of $\frac{\mu}{T}$ happens at
\be
m=(\frac{3 q^4}{4})^{\frac{1}{3}}
\ee
where $\frac{\mu}{T}=1.1107$. This is in fact the critical point -as the behavior of thermodynamic quantities near this point implies- where these two branches merge and is shown as the black point in figure \ref{critical}, left. 
The question we asked in the beginning of this section can be addressed by checking the behavior of the response in the field theory to the time-dependent source. As discussed earlier in the paper, this is introduced by time-dependent scalar field in the bulk or  Vaidya background which produces time-dependent temperature and chemical potential in field theory. We will show that the response in field theory, although a one-point function, knows about the critical point. Following the same calculation done in the previous sections we have plotted the rescaled equilibration time with respect to $\frac{\mu}{T}$ for two different cases of injecting energy in figure \ref{critical}, right. The magenta dashed line in this graph shows the critical point where $\frac{\mu}{T}$ gets its maximum value. The stable(unstable) branches are presented by blue(red) points. A very interesting observation is that for each value of $\frac{\mu}{T}$ the rescaled equilibration time for the stable(unstable) branch is smaller(larger). It can be shown that for other cases of energy injection with different values of $\beta_{\phi}$ or $\beta_{BH}$ the same conclusion can be made. Therefore one can conclude if we are having the information only about the temperature and chemical potential in a strongly coupled gauge theory, the rescaled equilibration time calculated using the bulk gravity dual can distinguish between stable and unstable branches. But we should note that in the unstable case we have fixed the system by hand in order to obtain the equilibration time and in reality an unstable solution can give rise to different physics. 

\begin{figure}
\includegraphics[width=80mm]{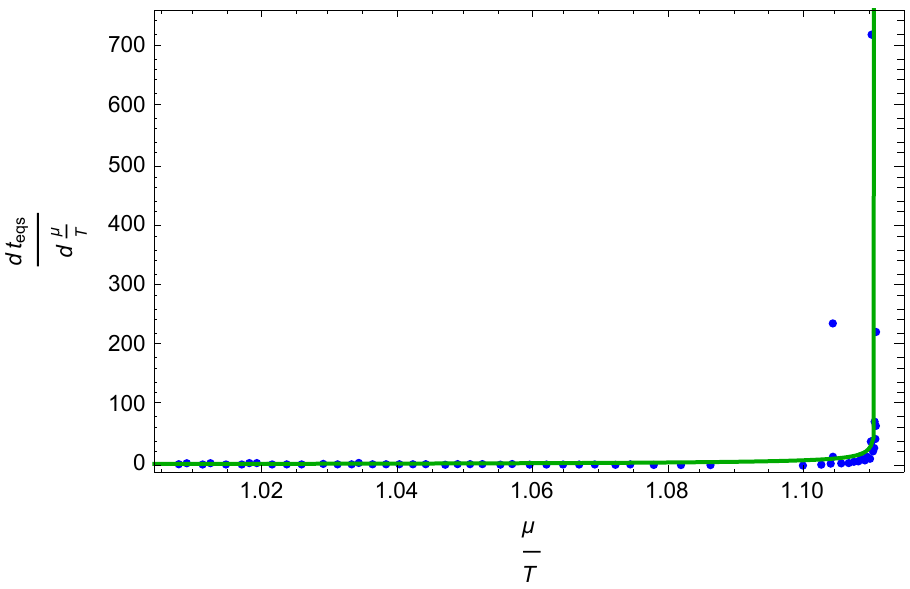}
\includegraphics[width=80mm]{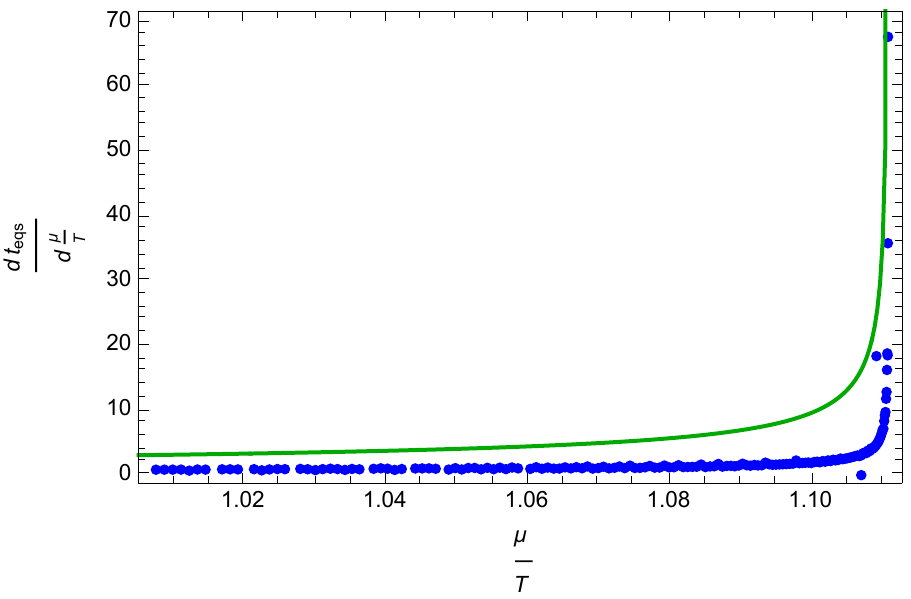}
\caption{$\frac{d t_{eqs}}{d \frac{\mu}{T}}$ with respect to $\frac{\mu}{T}$. Left: The scalar source has $\beta_{\phi}=0.2$. The green curve is the function, $(\frac{\pi}{2 \sqrt{2}}-\frac{\mu}{T})^{-\theta}$, fitted with the data with $\theta=0.489682$. Right: The scalar source has $\beta_{\phi}=5$. The green curve is the same function fitted with the data.} \label{exponent}
\end{figure}

If we look more closely to the plot near the critical point in figure \ref{critical}, right, we see that at the critical point, the magenta dashed line, the slope of the figure approaches infinity though the rescaled equilibration time is finite there. This suggests that the slope of the plot for the points near the critical point can be fitted with a function of the form $(\frac{\pi}{2 \sqrt{2}}-\frac{\mu}{T})^{-\theta}$ where $\theta$ is a positive number. In order to check we have plotted the slope in figure \ref{exponent} in the approximation where we have defined the slope as 
\be
\frac{d t_{eqs}}{d \frac{\mu}{T}} (i) = \frac{t_{eqs} (i+1) - t_{eqs} (i)}{\frac{\mu}{T} (i+1) - \frac{\mu}{T} (i)},
\label{slope}
\ee
where $i$ represents the ith point in the corresponding data points and $t_{eqs}$ is the rescaled equilibration time, vertical axes in figure \ref{critical}, right. In figure \ref{exponent} we have used the data points of figure \ref{critical}, right, in the stable branches. The figure \ref{exponent}, left(right), corresponds to the slope obtained from the top(bottom) plot in \ref{critical}, right. Interestingly, as shown in the figures \ref{exponent}, the slope data points can be fitted with the function
\be
\frac{d t_{eqs}}{d \frac{\mu}{T}} = (\frac{\pi}{2 \sqrt{2}}-\frac{\mu}{T})^{- \theta}, 
\ee
shown with the green curves in the figures. It seems that the value of $\theta$ depends on the quench being slow or fast. For the fast quench given by $\beta_{\phi}=0.2$ we obtain $\theta=0.489682$ which is very close to $0.5$, figure \ref{exponent}, left, and for slow quench $\beta_{\phi}=5$ we get $\theta=0.33901$, figure \ref{exponent}. If we compare our result here with the literature we can see the value $\theta=0.5$ is in fact what is called the dynamical critical exponent obtained from the behavior of scalar quasi-normal modes near the critical point in \cite{Finazzo:2016psx}. It is very appealing that we obtain the same result (with very good approximation) in the response of the system to a time-dependent scalar field in the probe limit. Therefore one can conclude that the response in field theory to a time-dependent source in the fast quench limit, where the source corresponds to a scalar field in the probe limit in the gravity dual, gives the dynamical critical exponent in field theory. In order to clarify more on the result we have obtained, we should point out that $\Delta \frac{\mu}{T}$ is between $10^{-2}$ and $10^{-6}$ and the value of $\theta$ for each subset of data points is reported in table \ref{list}. For each subset in this table the number of data points are chosen between 20 to 50.     

\begin{table}[ht]
\caption{data points of $min < \frac{d t_{eqs}}{d \frac{\mu}{T}} < max$ fitted with $(\frac{\pi}{2 \sqrt{2}}-\frac{\mu}{T})^{- \theta}$}
\vspace{3 mm}
\centering
\begin{tabular}{c c c c c c c c c}
\hline\hline
~~$ sets $ ~~   &~~$ set1 $ ~~   &   ~~ $ set2 $ ~~ & ~~ $ set3 $~~ & ~~$ set4 $ ~~   &   ~~ $ set5 $ ~~ & ~~ $ set6 $~~ & ~~$ set7 $ ~~   &   ~~ $ set8 $ ~~  \\[0.5ex]
\hline
$min$ & 0.144 & 1.108 & 1.309 & 1.604 & 2.015 & 2.501 & 3.010 & 4.14\\
$max$ & 1.100 &  1.295  & 1.597 & 1.998 & 2.499 & 2.989 & 3.959 & 721.229\\
$\theta$ & -1.853 &  0.084 & 0.233 & 0.230 & 0.348 & 0.462 & 0.422 & 0.490\\
\hline
\end{tabular}\\[1ex]
\label{list}
\end{table}

\section*{Acknowledgement}
M. A. would like to thank School of Physics of Institute for research in fundamental sciences (IPM) for the research facilities and environment.


\begin{thebibliography}{99}

\bibitem{Hubeny:2010ry} 
  V.~E.~Hubeny and M.~Rangamani,
  ``A Holographic view on physics out of equilibrium,''
  Adv.\ High Energy Phys.\  {\bf 2010}, 297916 (2010)
  doi:10.1155/2010/297916
  [arXiv:1006.3675 [hep-th]].
  
\bibitem{CasalderreySolana:2011us} 
  J.~Casalderrey-Solana, H.~Liu, D.~Mateos, K.~Rajagopal and U.~A.~Wiedemann,
  ``Gauge/String Duality, Hot QCD and Heavy Ion Collisions,''
  book:Gauge/String Duality, Hot QCD and Heavy Ion Collisions. Cambridge, UK: Cambridge University Press, 2014
  [arXiv:1101.0618 [hep-th]].

\bibitem{Maldacena:1997re}
  J.~M.~Maldacena,
  ``The Large N limit of superconformal field theories and supergravity,''
  Int.\ J.\ Theor.\ Phys.\  {\bf 38}, 1113 (1999)
  [Adv.\ Theor.\ Math.\ Phys.\  {\bf 2}, 231 (1998)]
  [hep-th/9711200].

\bibitem{Chesler:2008hg}
  P.~M.~Chesler and L.~G.~Yaffe,
  ``Horizon formation and far-from-equilibrium isotropization in supersymmetric Yang-Mills plasma,''
  Phys.\ Rev.\ Lett.\  {\bf 102}, 211601 (2009)
  [arXiv:0812.2053 [hep-th]];
  H.~Ebrahim and M.~Headrick,
  ``Instantaneous Thermalization in Holographic Plasmas,''
  arXiv:1010.5443 [hep-th];
  V.~Balasubramanian {\it et al.},
  ``Holographic Thermalization,''
  Phys.\ Rev.\ D {\bf 84}, 026010 (2011)
  [arXiv:1103.2683 [hep-th]];
  P.~M.~Chesler and L.~G.~Yaffe,
  ``Boost invariant flow, black hole formation, and far-from-equilibrium dynamics in N = 4 supersymmetric Yang-Mills theory,''
  Phys.\ Rev.\ D {\bf 82}, 026006 (2010)
  [arXiv:0906.4426 [hep-th]];
  D.~Kaviani,
  ``D7-brane dynamics and thermalization in the Kuperstein-Sonnenschein model,''
  arXiv:1608.02380 [hep-th];
  D.~Kaviani,
  ``Dp-brane dynamics and thermalization in type IIB Ben Ami-Kuperstein-Sonnenschein models,''
  arXiv:1708.00326 [hep-th];
  M.~Ali-Akbari and F.~Charmchi,
  ``Holographic Equilibration under External Dynamical Electric Field,''
  arXiv:1612.09098 [hep-th];
  A.~Dey, S.~Mahapatra and T.~Sarkar,
  ``Holographic Thermalization with Weyl Corrections,''
  JHEP {\bf 1601}, 088 (2016)
  doi:10.1007/JHEP01(2016)088
  [arXiv:1510.00232 [hep-th]].

\bibitem{Buchel:2014gta}
  A.~Buchel, R.~C.~Myers and A.~van Niekerk,
  ``Nonlocal probes of thermalization in holographic quenches with spectral methods,''
  JHEP {\bf 1502}, 017 (2015)
  [arXiv:1410.6201 [hep-th]].

\bibitem{Heller:2013oxa}
  M.~P.~Heller, D.~Mateos, W.~van der Schee and M.~Triana,
  ``Holographic isotropization linearized,''
  JHEP {\bf 1309}, 026 (2013)
  [arXiv:1304.5172 [hep-th]].

\bibitem{Buchel:2012gw} 
  A.~Buchel, L.~Lehner and R.~C.~Myers,
  ``Thermal quenches in N=2* plasmas,''
  JHEP {\bf 1208}, 049 (2012)
  [arXiv:1206.6785 [hep-th]].

      \bibitem{Finazzo:2016psx} 
        S.~I.~Finazzo, R.~Rougemont, M.~Zaniboni, R.~Critelli and J.~Noronha,
        ``Critical behavior of non-hydrodynamic quasinormal modes in a strongly coupled plasma,''
        JHEP {\bf 1701}, 137 (2017)
        [arXiv:1610.01519 [hep-th]].
 
\bibitem{Zhang:2015dia} 
  S.~J.~Zhang and E.~Abdalla,
  ``Holographic Thermalization in Charged Dilaton Anti-de Sitter Spacetime,''
  Nucl.\ Phys.\ B {\bf 896}, 569 (2015)
  [arXiv:1503.07700 [hep-th]].
  
  \bibitem{link}
  O.~Aharony, S.~S.~Gubser, J.~M.~Maldacena, H.~Ooguri and Y.~Oz,
  ``Large N field theories, string theory and gravity,''
  Phys.\ Rept.\  {\bf 323}, 183 (2000)
  doi:10.1016/S0370-1573(99)00083-6
  [hep-th/9905111];\\
  http://physics.ipm.ac.ir/phd-courses/semester8/AdS-CFT-lecturenotes-2013.pdf

\bibitem{Shahkarami:2017fxc} 
  L.~Shahkarami, H.~Ebrahim, M.~Ali-Akbari and F.~Charmchi,
  ``Far-from-equilibrium initial conditions probed by a nonlocal observable,''
  Phys.\ Lett.\ B {\bf 773}, 91 (2017)
  [arXiv:1702.08482 [hep-th]].
  
  \bibitem{Ali-Akbari:2016sms} 
  M.~Ali-Akbari, F.~Charmchi, H.~Ebrahim and L.~Shahkarami,
  ``Various Time-Scales of Relaxation,''
  Phys.\ Rev.\ D {\bf 94}, no. 4, 046008 (2016)
  [arXiv:1602.07903 [hep-th]].
  
  
\bibitem{Galante:2012pv} 
  D.~Galante and M.~Schvellinger,
  ``Thermalization with a chemical potential from AdS spaces,''
  JHEP {\bf 1207}, 096 (2012)
  [arXiv:1205.1548 [hep-th]];
  E.~Caceres and A.~Kundu,
  ``Holographic Thermalization with Chemical Potential,''
  JHEP {\bf 1209}, 055 (2012)
  [arXiv:1205.2354 [hep-th]];
  X.~X.~Zeng, X.~M.~Liu and W.~B.~Liu,
  ``Holographic thermalization with a chemical potential in Gauss-Bonnet gravity,''
  JHEP {\bf 1403}, 031 (2014)
  [arXiv:1311.0718 [hep-th]];
  G.~Camilo, B.~Cuadros-Melgar and E.~Abdalla,
  ``Holographic thermalization with a chemical potential from Born-Infeld electrodynamics,''
  JHEP {\bf 1502}, 103 (2015)
  [arXiv:1412.3878 [hep-th]].
  
  \bibitem{DeWolfe:2010he} 
    O.~DeWolfe, S.~S.~Gubser and C.~Rosen,
    ``A holographic critical point,''
    Phys.\ Rev.\ D {\bf 83}, 086005 (2011)
    doi:10.1103/PhysRevD.83.086005
    [arXiv:1012.1864 [hep-th]];
      O.~DeWolfe, S.~S.~Gubser and C.~Rosen,
      ``Dynamic critical phenomena at a holographic critical point,''
      Phys.\ Rev.\ D {\bf 84}, 126014 (2011)
      [arXiv:1108.2029 [hep-th]].
 
\bibitem{Rougemont:2015wca} 
  R.~Rougemont, A.~Ficnar, S.~Finazzo and J.~Noronha,
  ``Energy loss, equilibration, and thermodynamics of a baryon rich strongly coupled quark-gluon plasma,''
  JHEP {\bf 1604}, 102 (2016)
  doi:10.1007/JHEP04(2016)102
  [arXiv:1507.06556 [hep-th]].
  
\bibitem{Critelli:2017euk} 
  R.~Critelli, R.~Rougemont and J.~Noronha,
  ``Homogeneous isotropization and equilibration of a strongly coupled plasma with a critical point,''
  JHEP {\bf 1712}, 029 (2017)
  doi:10.1007/JHEP12(2017)029
  [arXiv:1709.03131 [hep-th]];
  R.~Rougemont, R.~Critelli, J.~Noronha-Hostler, J.~Noronha and C.~Ratti,
  ``Dynamical versus equilibrium properties of the QCD phase transition: A holographic perspective,''
  Phys.\ Rev.\ D {\bf 96}, no. 1, 014032 (2017)
  doi:10.1103/PhysRevD.96.014032
  [arXiv:1704.05558 [hep-ph]];
  S.~Janiszewski and M.~Kaminski,
  ``Quasinormal modes of magnetic and electric black branes versus far from equilibrium anisotropic fluids,''
  Phys.\ Rev.\ D {\bf 93}, no. 2, 025006 (2016)
  doi:10.1103/PhysRevD.93.025006
  [arXiv:1508.06993 [hep-th]].
  
  
\end{thebibliography}
\end{document}